# Temporally staggered cropping co-benefits beneficial insects and pest control globally

Adrija Datta[1], Subramanian Sankaranarayanan[2], Udit Bhatia[1,3,4*]

[1]Department of Earth Sciences, Indian Institute of Technology, Gandhinagar, Gujarat, India-382355
[2]Department of Biological Sciences and Engineering, Indian Institute of Technology, Gandhinagar, Gujarat, India-382355
[3]Department of Computer Science and Engineering, Indian Institute of Technology, Gandhinagar, Gujarat, India-382355
[4]Department of Civil Engineering, Indian Institute of Technology, Gandhinagar, Gujarat, India-382355
*Correspondence to: bhatia.u@iitgn.ac.in

**Abstract:**

Reconciling increasing food production with biodiversity conservation is critical yet challenging, particularly given global declines in beneficial insects driven by monoculture intensification. Intercropping—the simultaneous or sequential cultivation of multiple crops—has been proposed as a viable strategy to enhance beneficial insect services and suppress pests, yet global evidence regarding optimal spatiotemporal intercropping configurations remains fragmented. Here, we synthesize results from 7,584 field experiments spanning six continents and 22 Köppen climate regions, evaluating effects of spatial (row, strip, mixed, agroforestry) and temporal (additive, replacement, relay) intercropping configurations on beneficial insect (predators, parasitoids, pollinators) abundance and pest suppression using the Management Efficiency Ratio (MER; log ratio of abundance in intercropping versus monoculture). Relay intercropping, characterized by temporally staggered planting, emerged as the universally optimal temporal configuration, substantially increasing predator (MER = 0.473) and parasitoid populations (MER = 0.512) and effectively suppressing pests (MER = –0.611) globally. At regional scales, identical spatiotemporal configurations simultaneously optimized beneficial insect predator abundance and pest suppression in 57% of regions, while other regions required distinct, insect-specific approaches. Our findings highlight relay intercropping as a globally generalizable solution, but underscore regional variation that calls for targeted policies to simultaneously secure food production and biodiversity conservation.

# 1. Introduction

The United Nations Sustainable Development Goals (SDGs) set ambitious targets to concurrently achieve agricultural productivity, biodiversity conservation, and ecosystem sustainability by 2030[1]. However, prevailing agricultural practices significantly hinder progress towards these interconnected goals[2,3]. Agriculture alone drives nearly 90% of global deforestation[4,5] and threatens over 17,000 species worldwide[6]. Intensive monoculture farming, heavily dependent on agrochemicals[7], has particularly contributed to alarming declines in insect biodiversity[8–10], with more than 40% of insect species experiencing global reductions[11,12]. These insects underpin essential ecosystem services, including pollination[10,13] (critical for 75% of global food crops[14]), natural pest suppression[15], and nutrient cycling[16]. Consequently, identifying scalable agricultural practices that enhance crop productivity by controlling pests without compromising beneficial insect diversity has become critical, but requires a detailed global understanding of ecological outcomes associated with such practices across diverse climatic and geographic contexts.

Crop diversification practices, notably intercropping—the simultaneous or sequential cultivation of multiple crops within the same agricultural area[17]—have emerged as a viable strategy[18,19] to enhance agroecosystem resilience and reduce reliance on chemical pesticides[20,21]. Previous syntheses highlight intercropping's capacity to enhance beneficial insect populations[22,23], effectively suppress pests[24], and improve soil health[25] by increasing plant diversity and microclimatic complexity. However, global analyses[26,27] typically aggregate results across diverse intercropping systems without explicitly distinguishing between spatial (Fig. 1a) configurations (row, strip, mixed, agroforestry) or temporal (Fig. 1b) designs (additive, replacement, relay), each of which uniquely influences different insect population dynamics and ecosystem functioning (for more details, see *Methods*). Furthermore, despite extensive local-scale evidence documenting[28–31] positive ecological outcomes from specific intercropping configurations, these findings are inherently site-specific and cannot directly inform global or regional decision-making. Consequently, the absence of a unified global assessment examining how distinct spatiotemporal intercropping configurations systematically influence multiple insect functional groups across diverse Köppen climate regions limits the development of broadly applicable and effective agricultural sustainability strategies.

To address this knowledge gap, we synthesized global data from 7,584 field experiments reported in 336 peer-reviewed studies spanning 22 Köppen climate regions[32] across six continents (Fig. 1c). Using monoculture systems as controls, we quantitatively assessed how distinct spatiotemporal intercropping configurations influence key insect functional groups, including pollinators, predators, parasitoids, and pests. Specifically, our synthesis addressed three research questions: (1) Which spatial (row, strip, mixed, agroforestry) and temporal (additive, replacement, relay) intercropping configurations most effectively enhance beneficial insects and suppress pests? (2) Do ecological benefits of these configurations remain consistent across different insect functional groups and diverse climate regions? (3) How do specific crop combinations modulate different insect functional responses in varied climates? We addressed these questions using meta-analytic methods and non-parametric statistical tests to quantify insect responses across multiple geographic and climatic contexts. Addressing these questions at a global scale enables assessment of different spatiotemporal intercropping configurations as a scalable sustainability practice. Our results thus provide critical insights into how region-specific intercropping strategies can be systematically optimized, with direct implications for informing agricultural policy, guiding biodiversity conservation efforts, and achieving practical alignment between crop productivity and ecological resilience worldwide.

## 2. Data

**2.1 Study Selection:**
A literature search was conducted on Google Scholar (last accessed in February 2025) using two search strings:
i) ["Intercrop" AND "Insect abundance"], and
ii) [(("insect" AND "pest") OR ("insect" AND "pollinator") OR ("insect" AND "predator") OR ("insect" AND "parasitoid")) AND "intercrop" AND "abundance"].
The first 1000 results from each search were screened. Each article was evaluated for relevance based on defined inclusion criteria. From the first search string, 298 articles met the criteria, while 38 articles qualified from the second. The review encompassed studies published between 1978 and January 30, 2025.

A total of 336 articles were selected based on a two-stage screening process. The initial screening applied the following inclusion criteria: 1) the article was written in English; 2) duplicate studies published under different titles were excluded; 3) full-text access was available either through open access or institutional access; and 4) the study was peer-reviewed.

This initial screening resulted in 819 articles. These were then evaluated against additional criteria to determine final eligibility: 5) opinion, information bulletins, reviews, and meta-analyses were excluded; 6) studies needed to compare monoculture (as a control) with intercropping (as a treatment); 7) articles had to assess insect taxonomic diversity metrics, specifically abundance; 8) at least one insect species had to be identified and included in the analysis; 9) studies had to follow the definition of intercropping as the simultaneous cultivation of crops in the same field; 10) only open-field experiments were considered i.e. greenhouse or laboratory studies were excluded; 11) studies involving the use of chemical or organic pesticides were not included; 12) research involving transgenic or non-edible crops was excluded; 13) modeling-based studies were not considered; 14) studies using multiple intercrop species were excluded to avoid confounding effects of interspecific competition on insects; 15) studies focusing on insects in stored grains were not included; 16) research involving border or cover crops instead of true intercrops was excluded; and 17) studies that presented data only in 3D graphs or in formats that could not be extracted were not considered.

The search resulted in 336 articles deemed suitable for inclusion in our meta-analysis, from which a total of 7,584 effect sizes were extracted ([Fig. S10](Fig. S10)). The large number of effect sizes is primarily due to many studies evaluating multiple insect taxonomic groups when comparing intercropping with monoculture systems.

**2.2 General information on the articles:**
'Title', 'Authors', 'DOI', and 'Year_of_publication' variables: These variables provide essential information for easily locating the articles included in our database. They comprise the full title of the article, the list of authors, the Digital Object Identifier (DOI), and the year in which the article was published.

**2.3 Experimental site and climate conditions:**
'Country', 'Study_site', 'Latitude', and 'Longitude' variables: These variables provide location-specific details of the experimental sites, including the country and specific site where the study was conducted, along with the latitude and longitude (in decimal degrees). Coordinates were obtained either directly from the articles or estimated using the site name via Google Maps (https://maps.google.com/). The 'Climate_zone' variable is classified based on the Köppen-Geiger climate classification system[32].

**2.4 Spatial intercropping configuration description:**
'Spatial_intercropping': This variable describes the spatial arrangement used for sowing both species in the intercropping system[33]. (Fig. 1a).
1. *Row intercropping*: Where the two plant species are grown in distinct, alternating rows[34].
2. *Strip intercropping*: Where multiple rows of one plant species are alternated with one or more rows of another species, forming strips in which each strip consists of more than a single row[35].
3. *Mixed intercropping*: Where the component crops are sown at the same time, either within the same row or in a mixed pattern without any distinct rows or strips[36].
4. *Agroforestry*: All the agroforestry systems included here were alley cropping systems[37]. Here in alley cropping system, rows of crops are planted between rows of trees or shrubs[38].

**2.5 Temporal intercropping configuration description:**
'Temporal_intercropping': This variable describes the temporal arrangement used to intercrop both species. (Fig. 1b):
1. *Additive design:* In the standard additive design, intercropping systems are created by combining the plant densities of the respective pure stands. Consequently, the total density in the intercropping system is higher than in the pure stands, while the density of each component remains the same as in its pure stand[39].
2. *Replacement (or substitutive) design*: In the standard replacement design, intercropping systems are established by substituting a specific number of plants from one component with an equal number of plants from the other component. As a result, the density of each component in the mixture is lower than in its pure stand, but the overall stand density remains the same as in each pure stand[39].

3.  *Relay design:* In a standard relay intercropping design, the second crop is planted into the standing first crop at a later growth stage, creating a temporal overlap between the two crops instead of simultaneous planting. Although both crops occupy the same field space for part of their growth periods, the total plant density changes over time, and the density of each crop during the overlap phase is typically lower than in their respective monocultures[40].

**2.6 Crop details and insect abundance variables:**

Since the intercropping experiments include two species, the demonstration is made for one species (called Crop_1) and is replicable for the second species (Crop_2).

*'Crop_1_Common_Name' and 'Crop_1_Scientific_Name':* These variables provide both the scientific and common names of each crop species. To ensure consistency and avoid confusion caused by multiple common names referring to the same species, scientific names were matched with their corresponding common names using the United States Department of Agriculture (USDA) Plants Database (http://plants.usda.gov/java/).

*'Crop_1_family':* This variable indicates the botanical family of each crop species. Family information was either obtained directly from the source article or updated using the World Crops database (https://world-crops.com/category/crops/).

*'Crop_1_C':* This variable indicates the crop's photosynthetic pathway, identifying whether it follows the C3 or C4 carbon fixation process.

*'Crop_1_type':* This variable defines the crop category, such as cereals, legumes, vegetables, oilseeds, spices, forage, flowers, fruits, or trees. Crop classifications were updated using the World Crops database (https://world-crops.com/category/crops/). Crops without a clearly defined use were categorized as 'others'.

*'Insect_scientific_name' and 'Insect_common_name':* These variables include both the scientific and common names of each insect species. While some articles report scientific names, others mention only common names. In cases where only common names are provided, we used the GBIF Python package, pygbif (https://techdocs.gbif.org/en/openapi/), to retrieve the corresponding scientific names. For articles that only specify the insect's order or family without identifying the species, both the common and scientific name fields are left blank.

*'Insect_order' and 'Insect_family':* These variables describe the order and family of each insect species, primarily extracted directly from the original articles. For articles that do not report order or family but include either the scientific or common name, we used the GBIF Python package, pygbif (https://techdocs.gbif.org/en/openapi/), to determine their taxonomic classification.

*'Count_type':* This variable indicates the method used to count insects in each study, with the count type extracted directly from the respective article. This variable also includes the unit in which insect abundance is reported.

*'Insect_role' and 'Insect_role_type':* The 'Insect_role' variable broadly categorizes each insect based on its impact on crops, indicating whether it is beneficial or a pest. The 'Insect_role_type' variable further classifies beneficial insects into specific functional groups, such as pollinators, parasitoids, or predators.

*'Abundance_Crop_1' and 'Abundance_intercropped':* These two variables indicate insect abundance in monocropping ('Abundance_Crop_1') and intercropping ('Abundance_intercropped') systems, respectively. The method used to quantify insect counts is specified in the 'Count_type' variable.

## 3. Methods

### 3.1 Data extraction and collection:

Data were extracted from tables, figures, or textual descriptions within the articles. Values presented in graphs were manually digitized using the WebPlotDigitizer tool (https://automeris.io/WebPlotDigitizer/). All extracted data were compiled into an Excel spreadsheet, a format commonly supported by data analysis tools and well-suited for ensuring interpretability in scientific research. Additionally, information was documented on crop combinations, spatial arrangements, and the functional group classification of the insect species reported in each study.

A single author carefully evaluated each publication to determine its suitability and the reliability of the data. In total, 107 publications (representing 32% of all entries) underwent two rounds of review, once in November and again in February, to minimize potential errors. Data accuracy was verified multiple times by revisiting the original source materials whenever necessary.

### 3.2 Calculation of Management Efficiency Ratio (MER):

We calculated the Management Efficiency Ratio (MER) to compare insect abundance in the treatment (intercropping) versus the control (monoculture) systems (Equation 1).

$$MER = ln\frac{Abundance\ in\ treatment\ (intercrop)}{Abundance\ in\ control\ (monocrop)} \quad (1)$$

An MER greater than 0 indicates that intercropping has a positive effect on insect abundance relative to monoculture. In contrast, an MER less than 0 suggests a negative effect of intercropping compared to monoculture.

### 3.3 Publication bias analysis:

Egger's regression test[41] was conducted to detect potential funnel plot asymmetry as an indicator of publication bias. To further examine robustness against small-study effects, we applied a limit meta-analysis, estimating the effect size as sampling error approaches zero.

### 3.4 Statistical Analyses:

Anticipating variation in effect sizes across studies, we used the Mann-Whitney U test[42] to compare two major independent groups. Statistical significance was determined at a threshold of $p < 0.05$. To assess the magnitude of difference between groups, independent of sample size, we applied Cliff's Delta test[43]. A value of 0 indicates complete overlap between groups, while values approaching -1 or +1 reflect a strong effect.

To further explore subgroup differences, we assessed statistical heterogeneity using the Kruskal-Wallis test[44] and Higgins & Thompson's I² statistic[45]. The I² value quantifies the degree of heterogeneity: low ($\leq 25\%$), moderate (25–75%), and substantial ($\geq 75\%$).

In addition to the overall group comparisons, we conducted Dunn's test[46] with Bonferroni correction[47] to perform pairwise comparisons among the different intercropping strategies. This post-hoc analysis allowed us to identify which specific groups differed significantly after detecting overall differences using the Kruskal-Wallis test. The Bonferroni correction is crucial in this context as it adjusts for multiple comparisons, thereby reducing the risk of Type I errors (false positives) when testing multiple group differences simultaneously. While Dunn's test identifies statistically significant differences between groups, it does not convey the magnitude of these differences. Therefore, we calculated Cliff's Delta for each pairwise comparison to provide a non-parametric estimate of effect size.

To further investigate whether different spatiotemporal intercropping configurations significantly differ from one another, we applied the Kruskal-Wallis test along with Leave-One-Out Cross-Validation (LOO-CV)[48]. The LOO-CV approach was used to confirm the robustness of the Kruskal-Wallis results, ensuring that the findings were not disproportionately influenced by any single study or data point.

### 3.5 Identification of best strategies for each climate zone and crop combination:

The intercropping pattern and its structural design were combined into a single descriptive label to uniquely represent each practice. Records with missing or incomplete information on either the intercropping strategy, its design, or its performance metric were removed to ensure data quality. For each climate zone and insect role category, the average performance of each intercropping practice was calculated. This allowed for summarizing how effective different strategies were across various climatic contexts. To identify the most suitable strategy for each insect group in each climate zone, the following criteria were applied: i) For insect pests, the strategy with the lowest average performance metric (lowest MER) was selected (indicating greater effectiveness in suppression). ii) For beneficial insects (such as pollinators and natural enemies), the strategy with the highest average value (highest MER) was considered optimal.

The same strategy was followed to identify the most effective crop combinations (monocrop-intercrop combination) for each insect group and climatic region.

We have also repeated the same methodology to identify the intercropping pattern design combination for each insect group on each continent.

### 4. Results

### 4.1 Global research highlights regional gaps in crop diversity and insect studies

Geospatial analysis of 336 peer-reviewed studies published between 1978 and 2025 revealed marked regional biases, with most studies concentrated in tropical savanna (Aw, 24.5%), temperate oceanic (Cfb, 9.9%), and humid subtropical (Cwa, 9.3%) climate zones. Asia emerged as the primary research hub (39.2% of studies), followed by North America (21.1%) and Africa (13.7%). Country-level analyses further identified India (10.3%), Costa Rica

(5.5%), and Iran (5.2%) as key study locations. Asian studies primarily originated from India, Iran, Indonesia, China, and Pakistan, whereas African research centered predominantly in Tanzania, Benin, Uganda, Egypt, and Nigeria (Fig. S1a, S1b).

Intercontinental differences were pronounced in crop diversity. Asian studies reported the greatest diversity with 160 crop species, notably cabbage (28 intercrop combinations) and maize (22 combinations). African research employed 100 distinct crops (Fig. 2a), predominantly maize (37 combinations) and okra (19 combinations). Comparatively, developed regions displayed lower crop diversity: Europe had 46 documented crops (cabbage most common with 9 combinations), and North America reported 50 crops (maize with 7 combinations) (Fig. S2).

Insect community composition exhibited distinct biogeographic patterns. Asia recorded the highest taxonomic diversity (16 orders), dominated by Hemiptera (26.5%) and Lepidoptera (24.6%) (Fig. 2b). African studies documented nine insect orders, predominantly Lepidoptera (35.6%). Coleoptera dominated insect communities in Europe (64.1%), North America (59.1%), and Oceania (64.7%), whereas Hymenoptera was the leading order in South America (45.6%).

Globally, Lepidoptera was the most frequently studied insect order (25.3%), followed closely by Coleoptera (24.9%) and Hemiptera (23.8%) (Fig. 2c). Additional notable orders included Hymenoptera (9.9%), Diptera (4.0%), Neuroptera (3.2%), and Thysanoptera (3.0%). Among climate regions, tropical savanna (Aw) zones harbored the greatest insect diversity (25.4%), predominantly Lepidoptera (41.3%), followed by arid steppe (BSh, 12.2%), similarly dominated by Lepidoptera (35.3%) (Fig. 2d, Fig. S3, Table S1).

## 4.2 Global intercropping studies predominantly reflect pest suppression objectives

We assessed insect responses to intercropping using the Management Efficiency Ratio (MER), calculated as the natural logarithm of insect abundance ratios in intercropped versus monoculture systems as control (see *Methods* for details). If MER is positive, it means insect abundance will increase in intercropping with respect to monoculture and vice versa. Globally, intercropping resulted in predominantly negative MER values (63.6% of cases), indicating generally lower insect abundance in intercropping relative to monocultures (Fig. 3a). The global mean MER was significantly negative ($-0.164 \pm 0.652$; $p < 0.0001$), underscoring an overall suppressive effect of intercropping on insect populations. This globally observed suppression primarily reflects the predominant research emphasis on pest control, closely aligned with agricultural yield protection goals. Pest-focused studies were substantially more prevalent (n = 4,138) than studies assessing beneficial insects (pollinators and natural enemies, n = 2,540; Fig. 3b). Although intercropping is highly effective for pest management, the broader potential of intercropping to support beneficial insects remains likely underestimated due to these research biases.

This global trend, however, varied markedly across continents (Fig. S4, Table S2). South America showed a marginal increase in insect abundance under intercropping (weighted mean MER = 0.078). In contrast, Africa exhibited the strongest insect population suppression (weighted mean MER = –0.303).

While the global trend clearly aligns with pest suppression objectives, examining responses of different insect functional groups can further clarify whether beneficial insects exhibit systematically different responses under intercropping.

### 4.3 Intercropping selectively enhances beneficial insects and suppresses pests

We quantified the differential effects of intercropping on beneficial insects versus pests by comparing their Management Efficiency Ratios (MER). Beneficial insects exhibited significantly higher MER values (mean = 0.292 ± 0.012 SE), indicating increased abundance under intercropping compared to monocultures. In contrast, pests showed significantly negative MER values (mean = –0.445 ± 0.008 SE), with significant heterogeneity across groups ( $p < 0.001$; Cliff's delta = 0.67; Fig. 3b).

Further analysis of beneficial insects by functional group—pollinators, predators, and parasitoids—revealed that intercropping differentially benefited these groups. Predators (MER = 0.276 ± 0.015) and parasitoids (MER = 0.303 ± 0.025) responded positively and significantly more strongly than pollinators (MER = 0.076 ± 0.059), suggesting natural enemies derive the greatest advantage from intercropping (Fig. 3c).

A Kruskal–Wallis test[44] confirmed significant variation among these functional groups ($p < 0.001$; Higgins & Thompson's $I^2$ = 99.85%; Fig. S5a). Post hoc Dunn's pairwise comparisons[46] (Bonferroni-corrected[47]) indicated predators (Cliff's delta $\Delta$ = 0.203, $p < 0.001$) and parasitoids ($\Delta$ = 0.218, $p < 0.001$) benefited significantly more than pollinators. The difference between predators and parasitoids was statistically significant but negligible in magnitude ($\Delta$ = 0.006, $p < 0.05$). Pests showed consistent and substantial negative responses relative to all beneficial groups (e.g., $\Delta$ = –0.659 vs. predators, $p < 0.001$), underscoring the selective enhancement of natural enemies and effective pest suppression under intercropping.

Given that beneficial insect and pest populations respond distinctly, identifying specific spatiotemporal intercropping designs responsible for these differential outcomes is essential for targeted agroecological interventions.

### 4.4 Effective insect management emerges from aligning spatial and temporal intercropping designs with ecological roles

We analyzed global data to identify which temporal (additive, replacement, relay) and spatial (mixed, row, strip, agroforestry) intercropping configurations best enhance beneficial insect abundance and suppress pests. Among temporal designs, relay intercropping showed the strongest positive effect on parasitoids (mean MER = 0.473 ± 0.094 SE) and predators (mean MER = 0.512 ± 0.067 SE) and the greatest suppression of pests (mean MER = –0.611 ± 0.045 SE; Fig. 3d). Kruskal–Wallis analyses confirmed significant differences across temporal designs for predators ($p < 0.001$), parasitoids ($p < 0.01$), and pests ($p < 0.001$). Pairwise comparisons (Dunn's tests with Bonferroni correction, Fig. S5b) revealed relay designs significantly outperformed additive ($p < 0.05$) and replacement ($p < 0.001$) designs for both predators and parasitoids. These findings suggest relay intercropping provides beneficial insects with temporally continuous habitats and resources, while simultaneously disrupting pest colonization dynamics.

Spatial intercropping patterns also significantly influenced insect functional groups. Row intercropping yielded the highest MER values for pollinators (mean MER = 0.284 ± 0.132 SE) and predators (mean MER = 0.382 ± 0.02 SE), potentially due to enhanced foraging efficiency and accessibility (Fig. 3e, Fig. S6). Parasitoids responded best to strip (mean MER = 0.374 ± 0.048 SE), likely reflecting improved habitat segregation and host availability. Mixed intercropping most effectively suppressed pests (mean MER = –0.533 ± 0.076 SE), suggesting that greater spatial heterogeneity disrupts pest colonization and reproduction.

Finally, considering combined spatial and temporal dimensions, specific configurations emerged as optimal for different insect groups (Fig. S7, Fig. S8). Strip-replacement intercropping provided the greatest enhancement of parasitoid abundance (mean MER = 0.647 ± 0.151 SE), while row-additive intercropping best supported pollinators (mean MER = 0.284 ± 0.132 SE). Predators benefited most strongly from row-replacement intercropping (mean MER = 1.022 ± 0.135 SE, $p < 0.001$), and pest suppression was maximized in strip-relay systems (mean MER = –0.934 ± 0.046 SE, $p < 0.001$). Collectively, these results demonstrate that optimal insect management through intercropping depends critically on aligning spatial and temporal crop configurations with the specific ecological roles and habitat requirements of targeted insect groups.

Finally, since ecological roles and habitat preferences underpin insect responses to intercropping, it remains crucial to determine whether optimal strategies generalize across climate zones or require region-specific tailoring.

## 4.5 Climate-specific intercropping and crop combinations simultaneously optimize beneficial insect abundance and pest suppression

Given the broad climatic variability of global croplands, we conducted a region-by-region comparison through mosaic analysis (Fig. 4a) across major Köppen climate zones to identify optimal intercropping configurations for enhancing beneficial insects and suppressing pests. Pollinator responses were excluded due to insufficient data. In regions with adequate information, we determined the intercropping configurations that maximized beneficial insect abundance and minimized pest populations.

Our analysis revealed strong consistency in optimal intercropping choices among insect groups within many climate regions. In approximately 57% of regions, predators and pests responded most favorably to the same intercropping configuration, highlighting substantial ecological compatibility (Fig. 4a). Notably, in nearly one-third (28.5%) of the climate regions studied, a single intercropping configuration simultaneously optimized abundance of pests, predators, and parasitoids, underscoring potential for integrated insect management. Among these broadly beneficial strategies, row-additive and strip-additive intercropping designs emerged most frequently.

Crop combinations also strongly influenced insect responses. Cereal-legume intercrops were most consistently effective, emerging as the optimal combination in 18% of climate regions and frequently benefiting predators and parasitoids while reducing pests (Fig. 4b). Overall, 28% of regions favored one of six specific crop combinations—including cereals-legumes, spices-vegetables, cereals-cereals, flowers-fruits, cereals-oilseeds, and forage-vegetables—for simultaneously increasing predator abundance and suppressing pests. These intercrop combinations likely provide complementary ecological resources such as nectar, alternative prey, or structural habitats favorable to beneficial insects and may emit compounds that deter pests. However, no single crop combination universally promoted parasitoid and predator abundance while also suppressing pests, indicating specialized insect-crop interactions across different agricultural systems.

### 4.6 Publication bias report:
Egger's test[41] (Fig. S9) for funnel plot asymmetry revealed no significant publication bias (z = 1.4486, p = 0.1474). Additionally, the limit estimate, representing the effect size as sampling error approaches zero, was even more negative (b = -0.1916; 95% CI: -0.2845 to -0.0987), reinforcing the robustness of the observed effect.

## 5. Discussion:

Agricultural landscapes worldwide face the dual challenge of increasing food production and conserving biodiversity, two goals traditionally viewed as competing rather than complementary[6,49]. Historically, research has predominantly emphasized pest suppression[18] or yield enhancement[15,16] in isolation, resulting in a fragmented understanding of the broader agroecological potential of intercropping. Our global synthesis of 7,584 field experiments spanning six continents and 22 Köppen climate regions offers empirical evidence that intercropping can simultaneously enhance beneficial insect populations and suppress pest abundance across diverse agroecosystems. Our results demonstrate that relay intercropping—characterized by temporally staggered planting—is the universally optimal temporal configuration, significantly increasing predator (mean MER = 0.512) and parasitoid (mean MER = 0.473) populations while substantially suppressing pest numbers (mean MER = –0.611). Mosaic analyses further revealed considerable regional consistency, with identical spatiotemporal configurations simultaneously optimizing natural enemy abundance and pest suppression in approximately 28.5% of climate regions. Nevertheless, notable deviations among insect functional groups and climate zones highlight important geographic variability, underscoring the necessity for region-specific intercropping strategies.

Despite these robust global patterns, several critical knowledge gaps remain. Notably, our synthesis identified limited global-scale data on pollinator responses to intercropping, despite pollinators' critical roles in crop productivity[50] and food security[51]. Furthermore, arid and boreal climates are significantly underrepresented in existing datasets, restricting comprehensive global generalizations. Importantly, the absence of integrated yield and profitability assessments alongside insect abundance metrics limits comprehensive evaluations of intercropping's multifunctional benefits. Addressing these limitations by explicitly focusing on pollinator populations, expanding research coverage in underrepresented climates, and incorporating economic and yield outcomes will strengthen future global syntheses and enhance practical agricultural relevance.

Our findings have important implications for agricultural policy and management. While several countries have integrated intercropping and agroforestry into formal frameworks, such as India's integrated intercropping and agroforestry initiatives of Council on Energy, Environment and Water (CEEW)[52] and the European Union's Common Agricultural Policy[53] (2023-27), policy attention remains limited to select configurations. Building on this policy momentum, the recent International Maize and Wheat Improvement Center (CIMMYT) Intercropping project (2023-2028) focuses exclusively on row-additive intercropping in South Asia[54], overlooking other potentially effective spatiotemporal strategies. Several countries have already integrated intercropping and agroforestry into formal policy frameworks. While our results highlight the ecological benefits of relay intercropping for enhancing beneficial insects and suppressing pests, it remains underrepresented in current implementation efforts. Relay intercropping also aligns with resource constraints common to smallholder farming and substantially reduces insecticide dependency[55], making it a strong candidate for inclusion in extension programs, incentive schemes, and training. Strengthening policy support for relay intercropping can help achieve sustainability goals, specifically the United Nations Sustainable Development Goals on food security (SDG 2: Zero Hunger) and terrestrial biodiversity conservation (SDG 15: Life on Land).

Considering projected climate impacts on global agriculture due to anthropogenic climate change[56], future research must evaluate intercropping strategies under evolving pest dynamics, shifting beneficial insect distributions, and increased climatic extremes. Predictive ecological modeling that incorporates projected climate scenarios can inform resilient cropping recommendations, ensuring intercropping systems maintain effectiveness under altered ecological conditions. Additionally, translating our global-scale insights into spatially explicit, region-specific recommendations will facilitate practical application, allowing policymakers and practitioners to target agricultural diversification strategies effectively across varied climates and agroecological contexts.

Ultimately, our global synthesis provides a robust ecological basis for multifunctional agriculture, highlighting strategic relay intercropping designs that simultaneously enhance beneficial insect populations and effectively suppress pests. These clearly quantified and globally consistent results represent actionable pathways to reconcile intensified food production with biodiversity conservation goals, traditionally viewed as conflicting objectives. By systematically addressing identified knowledge gaps, particularly around pollinator dynamics and yield assessments, future research can further refine and expand the applicability of intercropping strategies. Such targeted agricultural diversification can support both ecological resilience and agricultural productivity, directly contributing to global sustainability priorities and addressing pressing environmental and food security challenges worldwide.

**Author contributions:**
Conceptualization: AD. Data collection: AD, Analysis: AD. Visualization: AD, SS, UB. Writing – original draft: AD, SS, UB. Writing – review and editing: AD, SS, UB.

**Competing interests:**
The authors declare that they have no competing interests.

**Acknowledgements:**
This research was primarily funded by IIT Gandhinagar. The authors also thank the members of the Machine Intelligence and Resilience Laboratory at IIT Gandhinagar for their valuable discussions and constructive feedback on this manuscript.

**Data and materials availability:**
All data supporting the findings of this study are included in the main text and/or Supplementary Materials. Additional data are available at the following repository: https://doi.org/10.5281/zenodo.16897437 [57].

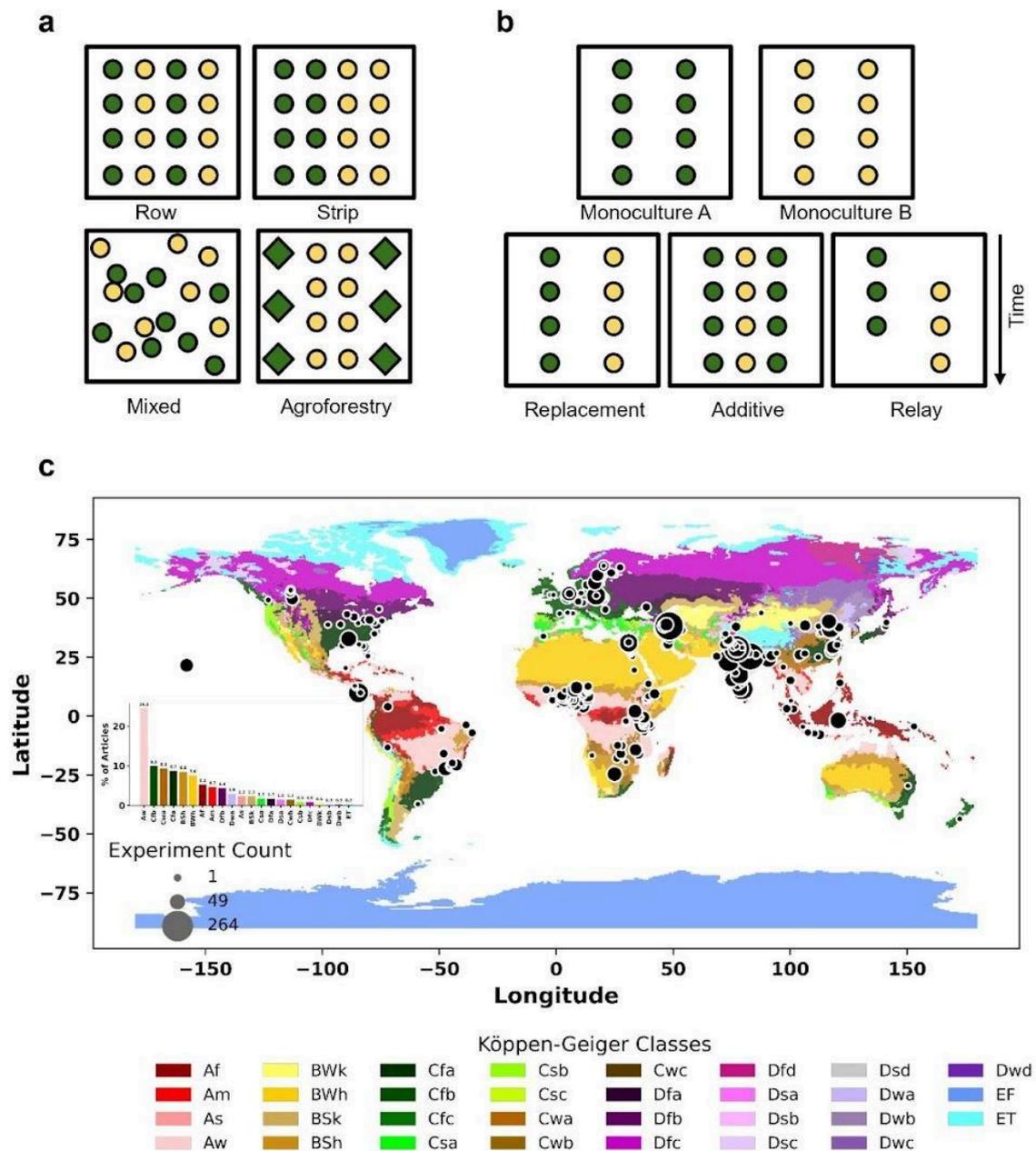

**Fig. 1| Spatial and temporal arrangements of crops in intercropping and geographical distribution of sites. a,** Spatial arrangements for pure stand crop A (●) and crop B (○) for row, strip, and mixed design, and for tree species (◆) in agroforestry. **b,** Temporal arrangements pure stand crop A (●) and crop B (○) for replacement, additive, and relay intercropping designs. **c,** Geographical distribution of sites and number of experiments included in the database. The size of black circles is proportional to the number of experiments recorded at each location. The Köppen-Geiger climate classification system was applied to associate each field site with a grid cell measuring 0.50° latitude by 0.50° longitude. This system categorizes climates into five primary zones, each designated by a letter: A for tropical, B for arid, C for temperate, D for continental, and E for polar regions. The bar plot shows the percentage of studies (%) in each climatic region.

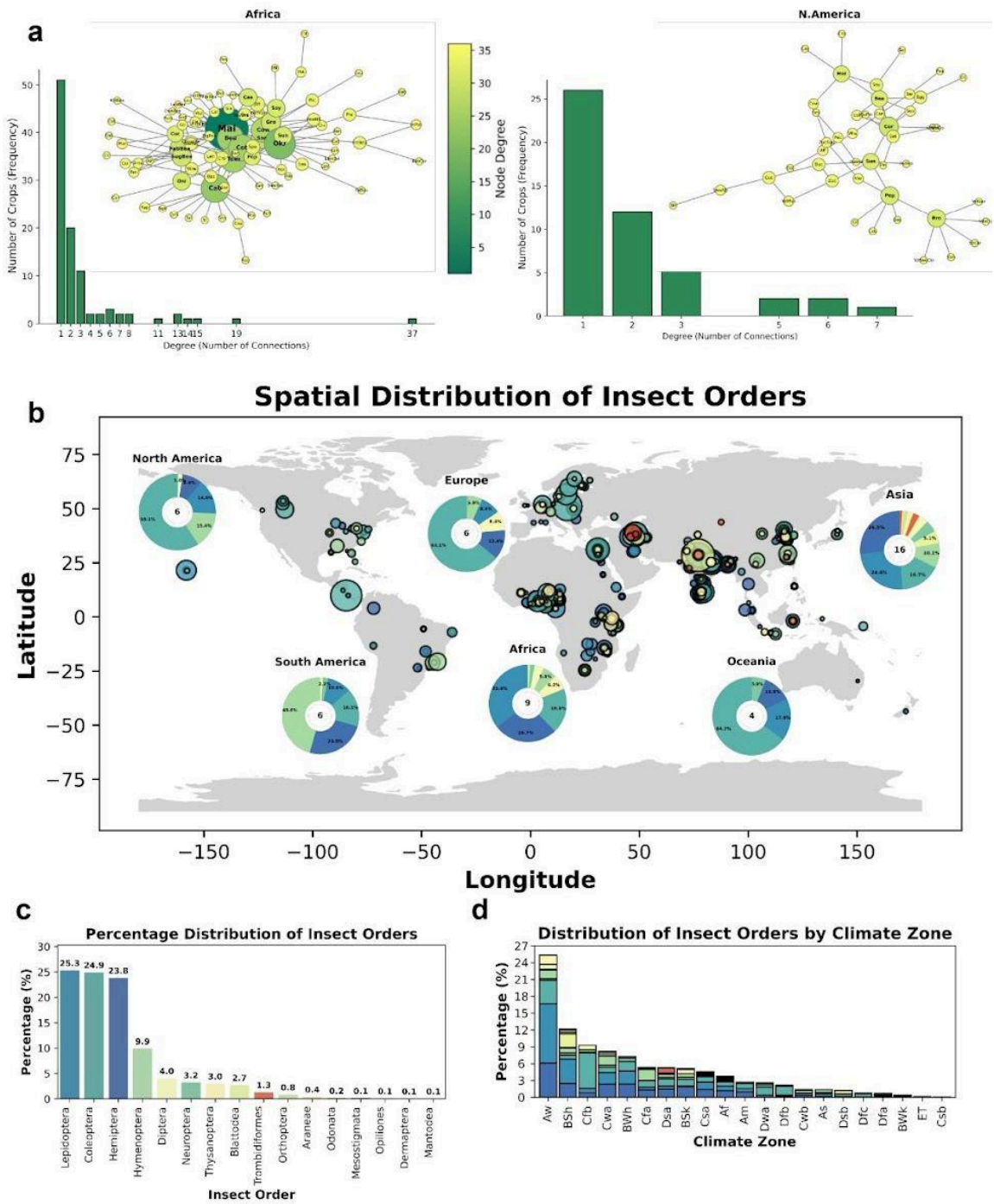

**Fig. 2| Patterns of crop interactions and insect order distributions in agroecosystems. a,** Crop interaction networks for Africa and North America, with nodes representing crops and edges showing intercropping pairs. Node size and color correspond to the number of intercropping connections, highlighting the degree (connection) of each crop within the network. Bar plots show degree distribution. **b,** Spatial distribution of insect orders, with dot size indicating the abundance of each order at each location. Donut charts show the continent-wise composition of insect orders. **c,** Global percentage distribution of reported insect orders. **d,** Distribution of insect orders across Köppen-Geiger climatic regions.

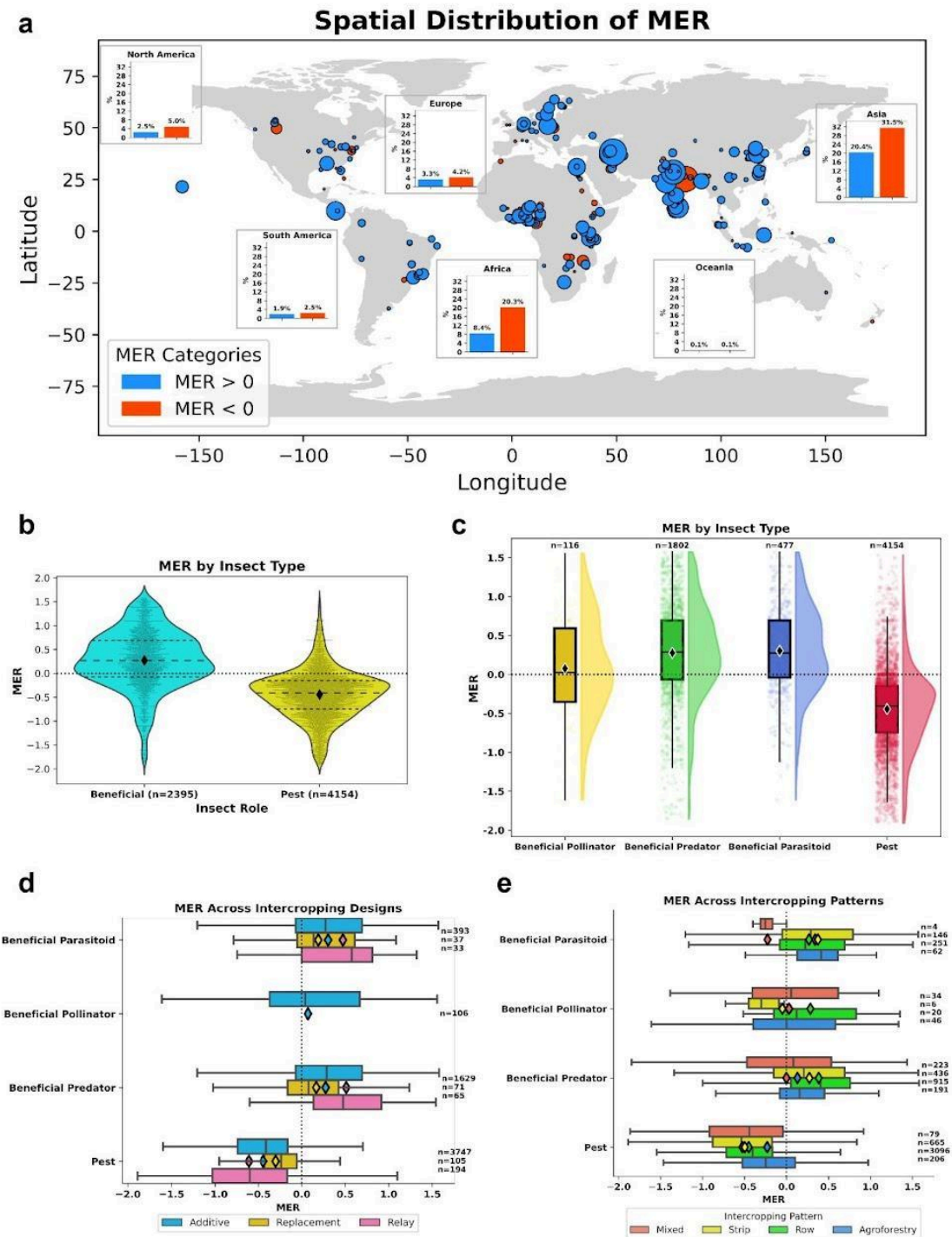

**Fig. 3| Impacts of spatiotemporal intercropping arrangements on insect functional groups. a,** Spatial distribution of Management Efficiency Ratio (MER). **b,** Effect of intercropping on beneficial insects and pests. **c,** Effect of intercropping on pollinators, parasitoids, and predators. **d,** Effect of temporal intercropping arrangements on insect functional groups. **e,** Effect of spatial intercropping arrangements on insect functional groups. The total number of individual effect sizes is indicated by "n." The diamond symbol illustrates the average effect size.

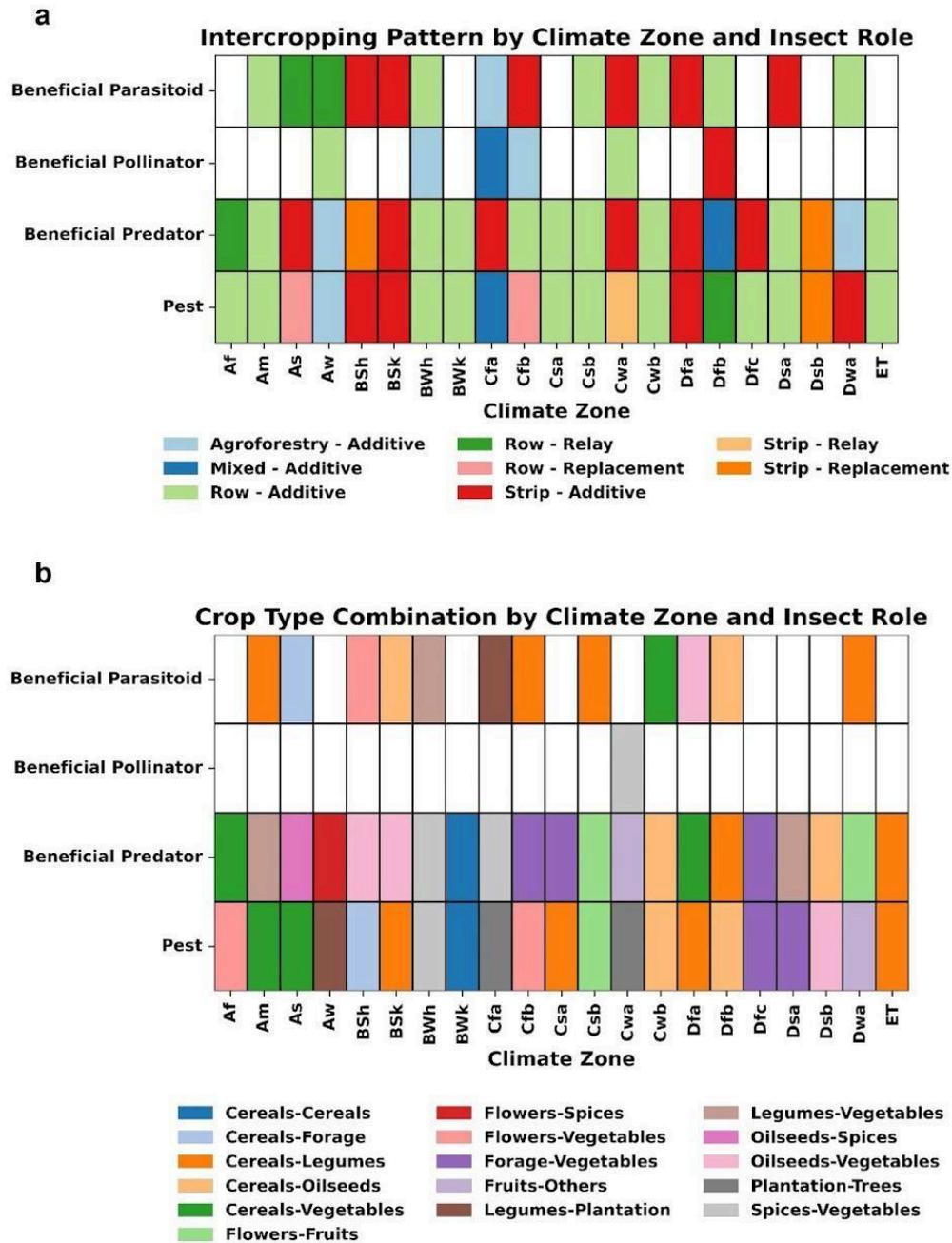

**Fig 4| Identification of region-specific spatiotemporal intercropping arrangements for different functional group management: a,** Mosaic map showing the most effective spatiotemporal intercropping arrangements for enhancing beneficial insect abundance and suppressing pest populations across Köppen-Geiger climate regions for each insect functional group. **b,** Mosaic map indicating the best crop combinations for enhancing beneficial insect abundance and suppressing pest populations for each Köppen-Geiger climate region and insect functional group. White areas represent either no data or only one persistent intercropping arrangement or crop combination.